\documentclass{aastex62}
\usepackage{xspace}
\usepackage{upgreek}
\usepackage{amsmath,amssymb}
\usepackage{tablefootnote}
\usepackage{hyperref}
\graphicspath{{./}{figures/}}

\newcommand{\firstlong}{FIRST~J141918.9+394036\xspace}
\newcommand{\first}{FIRST~J1419+3940\xspace}

\shorttitle{}
\shortauthors{Nimmo et al.}

\begin{document}

\title{No Radio Bursts Detected from \firstlong in Green Bank Telescope Observations}

\correspondingauthor{Kenzie Nimmo}
\email{k.nimmo@uva.nl}

\author[0000-0003-0510-0740]{Kenzie Nimmo}
\affil{ASTRON, Netherlands Institute for Radio Astronomy, Oude Hoogeveensedijk 4, 7991~PD Dwingeloo, The Netherlands}
\affil{Anton Pannekoek Institute for Astronomy, University of Amsterdam, Science Park 904, 1098 XH Amsterdam, The Netherlands}

\author[0000-0002-8604-106X]{Vishal Gajjar}
\affil{Department of Astronomy, University of California Berkeley, Berkeley, CA 94720, USA}

\author[0000-0003-2317-1446]{Jason~W.~T.~Hessels}
\affil{ASTRON, Netherlands Institute for Radio Astronomy, Oude Hoogeveensedijk 4, 7991~PD Dwingeloo, The Netherlands}
\affil{Anton Pannekoek Institute for Astronomy, University of Amsterdam, Science Park 904, 1098 XH Amsterdam, The Netherlands}

\author[0000-0002-4119-9963]{Casey~J.~Law}
\affil{Department of Astronomy and Owens Valley Radio Observatory, California Institute of Technology, Pasadena, California 91125, USA}

\author[0000-0001-5229-7430]{Ryan~S.~Lynch}
\affil{Green Bank Observatory, PO Box 2, Green Bank, WV 24944, USA}
\affil{Center for Gravitational Waves and Cosmology, West Virginia University, Morgantown, WV 26506, USA}

\author[0000-0001-6716-6126]{Andrew~D.~Seymour}
\affil{Green Bank Observatory, PO Box 2, Green Bank, WV 24944, USA}

\author[0000-0002-3775-8291]{Laura~G.~Spitler}
\affil{Max-Planck-Institut f\"ur Radioastronomie, Auf dem H\"ugel 69, D-53121 Bonn, Germany}






\section*{Introduction} \label{sec:intro}

Precise localisation of the first-known repeating fast radio burst source, FRB~121102 \citep{spitler2016, chatterjee2017}, led to its association with a star-forming region inside a low-metallicity dwarf host galaxy \citep{tendulkar2017}. This host environment is similar to that typically associated with long gamma-ray bursts (GRB) and superluminous supernovae, potentially linking these astrophysical phenomena \citep{metzger2017}. In addition, the bursting source is found to be spatially coincident with a compact ($<0.7\ \mathrm{pc}$; \citealt{marcote2017}), persistent radio source \citep{chatterjee2017}. \cite{ofek2017} identified similar radio sources in the Very Large Array FIRST survey \citep{becker1995}. One of these sources, \firstlong (hereafter \first), was identified as a radio transient decaying in brightness by a factor of $\sim 50$ over several decades \citep{law2018}. Very-long-baseline radio interferometric observations support the theory that \first is the afterglow of a long GRB, based on the inferred physical size of the emission region ($1.6\pm 0.3\ \mathrm{pc}$; \citealt{marcote2019}).

\first and FRB~121102's persistent radio sources have similar properties and host galaxy type. Although \first is declining in brightness, its peak luminosity ($\nu L_{\nu}> 3\times 10^{38}\ \mathrm{erg\ s^{-1}}$ at $1.4$\,GHz; \citealt{law2018}) is comparable to the mean luminosity of FRB~121102's persistent radio source ($\nu L_{\nu}\approx 3\times 10^{38}\ \mathrm{erg\ s^{-1}}$ at $1.7$\,GHz; \citealt{chatterjee2017}). Possibly, their physical nature could be similar, and \first could contain a source capable of producing millisecond-duration radio bursts. Above $\sim 1.4\ \mathrm{GHz}$, \first is observed to have an optically-thin synchrotron spectrum \citep{law2018}. This, combined with the relatively close proximity of \first ($87\ \mathrm{Mpc}$, about an order-of-magnitude closer than FRB~121102; \citealt{law2018}, \citealt{tendulkar2017}), indicates that it should be possible to detect much lower energy bursts than those observed from FRB~121102, if \first is producing FRBs. \cite{marcote2019} reported the non-detection of bursts from \first during $4.3\ \mathrm{h}$ of observations with the $100$-$\mathrm{m}$ Effelsberg telescope at $1.7$\,GHz.  Here, we report the non-detection of bursts from \first using the $110$-$\mathrm{m}$ Green Bank Telescope (GBT).

\section*{Observations and Analysis}

Table~\ref{tab:obs} summarises the observations.
We observed \first for a total duration of $3.1\ \mathrm{h}$ using the GBT and the Breakthrough Listen backend \citep{macmahon2018} on MJDs 58519 and 58529 --- at both S-band ($1.73$--$2.6$\,GHz) and C-band ($3.95$--$8.0$\,GHz).  The time and frequency resolutions were $349.5\ \mathrm{\upmu s}$ and $0.366\ \mathrm{MHz}$, respectively. In addition to the target scans, both noise diode and test pulsar (PSR~B1508+55) scans were taken. 

\begin{deluxetable*}{ccccccc}
    \tablecaption{Summary of observations and fluence upper limits.
    \label{tab:obs}}
    \tablehead{
       \colhead{Scan start time$^{1}$}  & \colhead{Frequency range} & \colhead{Duration} & \colhead{T$_\text{sys}$ + T$_\text{bg}\ ^{2}$} & \colhead{Gain} & \colhead{Fluence limit$^{3}$} \\
        \colhead{(MJD)} & \colhead{(${\rm GHz}$)}  & \colhead{(min)} & \colhead{(${\rm K}$)} & \colhead{(${\rm K/Jy}$)} & \colhead{(${\rm Jy\ ms}$)}
    }
    \startdata
        $58519.4457$ & $3.95$--$8.0$ & $30.0$ & $28$ & $1.85$ & $0.05$\\
        $58519.4667$ & $3.95$--$8.0$ & $30.0$ & $28$ & $1.85$ & $0.05$\\
        $58519.5059$ & $1.73$--$2.6$ & $6.7$ & $25$ & $1.9$ & $0.1$\\
        $58529.1809$ & $3.95$--$8.0$ & $30.0$ & $28$ & $1.85$ & $0.05$\\
        $58529.2019$ & $3.95$--$8.0$ & $30.0$ & $28$ & $1.85$ & $0.05$\\
        $58529.2305$ & $1.73$--$2.6$ & $30.0$ & $25$ & $1.9$ & $0.1$\\
        $58529.2514$ & $1.73$--$2.6$ & $28.0$ & $25$ & $1.9$ & $0.1$\\
    \enddata
    \tablenotetext{1}{Topocentric.}
    \tablenotetext{2}{System temperature (T$_\text{sys}$) are for typical GBT performance: \url{http://www.gb.nrao.edu/~fghigo/gbtdoc/perform.html}. Background temperature (T$_\text{bg}$) is a combination of the sky temperature (negligible in this case, using the $408\ \mathrm{MHz}$ all-sky map \citep{remazeilles2015} and extrapolating to our observing frequencies using a spectral index of $-2.7$ \citep{reich1988}) and the cosmic microwave background $\sim 3\ \mathrm{K}$ \citep{mather1994}.}
    \tablenotetext{3}{Calculated following \cite{cordes2003}, assuming a $1$-ms-wide burst with DM~$= 300\ \mathrm{pc\ cm^{-3}}$, using the temperature and gain values listed, with a signal-to-noise detection threshold of $10$.}
\end{deluxetable*}

We searched for bursts using {\tt PRESTO}\footnote{https://github.com/scottransom/presto} \citep{ransom2001}. We identified and masked radio frequency interference (RFI) using {\tt PRESTO}'s {\tt rfifind} and dedispersed using {\tt prepdata} to create timeseries with trial dispersion measures (DM) of $0-1000\mathrm{\ pc\ cm^{-3}}$. As is discussed in \cite{marcote2019}, the expected DM towards \first is $< 170\ \mathrm{pc\ cm^{-3}}$, ignoring any contribution from the host galaxy. If we assume the host contribution is comparable to that of FRB~121102, then the expected DM is $\sim 400\ \mathrm{pc\ cm^{-3}}$.   We then searched for single pulses above a $6\sigma$ threshold in the dedispersed time series using {\tt single\_pulse\_search.py}. The single pulses due to RFI were filtered using an automated classifier \citep{michilli2018c}.  Our search was sensitive to bursts with widths between $\sim0.5 \mathrm{\ ms}$ and $34.95\ \mathrm{ms}$. The identified candidates were all deemed to be non-astrophysical after inspecting their dynamic spectra by eye. This analysis strategy was verified by performing a blind search for the test pulsar PSR~B1508+55.

\section*{Results and Discussion}

In this search, we were sensitive to $1$-$\mathrm{ms}$-wide bursts from \first exceeding the fluence limits shown in Table \ref{tab:obs}, assuming DM $\sim 300\ \mathrm{pc\ cm^{-3}}$. Considering the weakest bursts observed from FRB~121102 ($0.02\ \mathrm{Jy\ ms}$; \citealt{gajjar2018}) and scaling to the luminosity distance of \first ($87\ \mathrm{Mpc}$; \citealt{law2018}), we find the corresponding fluence to be $2.5\ \mathrm{Jy\ ms}$, well exceeding our detection threshold.   We found no astrophysical bursts in these observations. If we assume there is a source associated with \first that is producing FRBs, the lack of detection could indicate a quiescent state, as is often observed for FRB~121102 \citep[e.g.][]{gajjar2018}. Alternatively, the bursts could be beamed away from our line-of-sight. It is also possible that \first does not contain a source capable of producing FRBs. Future searches are important to constrain the possible presence of an FRB-emitting source.


\begin{thebibliography}{}

\bibitem[Becker et~al. (1995)]{becker1995}
{Becker}, R.~H., {White}, R.~L. \& {Helfand}, D.~J. 1995, \apj, 450, 559 

\bibitem[Chatterjee et~al. (2017)]{chatterjee2017}
{Chatterjee}, S.,  {Law}, C.~J., {Wharton}, R.~S., et~al. 2017, \nat, 541, 58

\bibitem[Cordes \& McLaughlin (2003)]{cordes2003} Cordes, J.~M. \& McLaughlin, M.~A.\ 2003, \apj, 596, 1142 

\bibitem[Gajjar et al. (2018)]{gajjar2018} 
Gajjar, V., Siemion, A.~P.~V., Price, D.~C. et al.\ 2018, \apj, 863, 2

\bibitem[Law et al.(2018)]{law2018} 
Law, C.~J., Gaensler, B.~M., Metzger, B.~D. et al.\ 2018, \apj, 866, L22 

\bibitem[MacMahon et al. (2018)]{macmahon2018} 
MacMahon, D.~H.~E., Price, D.~C., Lebofsky, M. et al.\ 2018, \pasp, 130, 986 

\bibitem[Marcote et al. (2019)]{marcote2019}
Marcote, B., Nimmo, K., Salafia, O.~S. et al.\ 2019, \apjl, 876, L14

\bibitem[Marcote et al. (2017)]{marcote2017} 
Marcote, B., Paragi, Z., Hessels, J.~W.~T. et al.\ 2017, \apjl, 834, L8

\bibitem[Mather et al. (1994)]{mather1994}
{Mather}, J.~C., {Cheng}, E.~S., {Cottingham}, D.~A. et al.\ 1994, \apj, 420, 439

\bibitem[Metzger et al. (2017)]{metzger2017}
{Metzger}, B.~D.,  {Berger}, E. \&   {Margalit}, B.  2017, \apj, 841, 14

\bibitem[Michilli \& Hessels (2018)]{michilli2018c}
{Michilli}, D. and {Hessels}, J.~W.~T.\ 2018, SpS: Single-pulse Searcher

\bibitem[Ofek (2017)]{ofek2017} {Ofek}, E.~O. 2017, \apj, 846, 44 

\bibitem[Ransom (2001)]{ransom2001} 
{Ransom}, S.~M. 2001, PhD thesis, Harvard University

\bibitem[Reich \& Reich (1988)]{reich1988} Reich, P. \& Reich, W.\ 1988, Astronomy \& Astrophysics, 74, 7

\bibitem[Remazeilles et al. (2015)]{remazeilles2015} {Remazeilles}, M., Dickinson, C., Banday, A.~J., et al.\ 2015, \mnras, 451, 4311

\bibitem[Spitler et~al.(2016)]{spitler2016}
{Spitler}, L.~G., {Scholz}, P., {Hessels}, J.~W.~T., et~al. 2016, \nat, 531, 202

\bibitem[Tendulkar et~al. (2017)]{tendulkar2017}
{Tendulkar} S.~P., {Bassa}, C.~G., {Cordes}, J.~M.,  et~al. 2017, \apjl, 834, L7

\end{thebibliography}

\end{document}